# Anti-site defect-induced disorder in compensated topological magnet $MnBi_{2-x}Sb_xTe_4$

Felix Lüpke[1,2,3,‡], Marek Kolmer[1,4‡], Jiaqiang Yan[5], Hao Chang[1,6], Paolo Vilmercati[6,7], Hanno H. Weitering[6,7], Wonhee Ko[1,6,7*], An-Ping Li[1,6*]

[1]*Center for Nanophase Materials Sciences, Oak Ridge National Laboratory, Oak Ridge, Tennessee 37831, USA*

[2]*Department of Materials Science and Engineering, University of Tennessee, Knoxville, Tennessee 37916, USA*

[3]*Peter Grünberg Institut (PGI-3), Forschungszentrum Jülich, 52425 Jülich, Germany*

[4]*Ames National Laboratory, U.S. Department of Energy, Ames, Iowa 50011, USA*

[5]*Materials Science and Technology Division, Oak Ridge National Laboratory, Oak Ridge, Tennessee 37831, USA*

[6]*Department of Physics and Astronomy, The University of Tennessee, Knoxville, Tennessee 37996, USA*

[7]*Institute for Advanced Materials and Manufacturing at the University of Tennessee, 2641 Osprey Vista Way, Knoxville, Tennessee 37920, USA*

‡These authors contributed equally.

[*]Corresponding Authors: wko@utk.edu and apli@ornl.gov

## Abstract

The gapped Dirac-like surface states of compensated magnetic topological insulator $MnBi_{2-x}Sb_xTe_4$ (MBST) are a promising host for exotic quantum phenomena such as the quantum anomalous Hall effect and axion insulating state. However, it has become clear that atomic defects undermine the stabilization of such quantum phases as they lead to spatial variations in the surface state gap and doping levels. The large number of possible defect configurations in MBST make studying the influence of individual defects virtually impossible. Here, we present a statistical analysis of the nanoscale effect of defects in MBST with $x = 0.64$, by scanning tunnelling microscopy/spectroscopy. We identify $(Bi,Sb)_{Mn}$ anti-site defects to be the main source of the observed doping fluctuations, leading towards the formation of nanoscale charge puddles and effectively closing the transport gap. Our findings will guide further optimization of this material system via defect engineering, to enable exploitation of its promising properties.

## Introduction

Three-dimensional topological insulators such as $Bi_2Te_3$ and $Sb_2Te_3$ are established to have a topologically protected Dirac-like surface state with linear dispersion[1, 2]. Infusing these materials with magnetism breaks the time-reversal symmetry and opens a magnetic exchange gap at the Dirac point that gives rise to a plethora of exotic quantum states, such as the quantum anomalous Hall (QAH) effect, which can be exploited in quantum electronic devices[3]. The experimental signature of the QAH state has firstly been observed experimentally in chromium or vanadium-doped topological insulators of the $Bi_{2-x}Sb_xTe_3$ (BST) class[4-7]. In these compounds, the random distribution of a few percent of magnetic dopants opens a magnetic exchange gap, while Sb substitution allows tuning the charge carrier density so that the Fermi energy $E_F$ is located in vicinity to the exchange gap[8, 9]. However, the nature of the randomly distributed magnetic dopants in these compounds leads to variations in the exchange gap size as well as the doping level throughout the material, both of comparable magnitude[8, 9]. In transport experiments, these variations result in the formation of charge puddles, which inherently limit the device performance[9]. As a result, the QAH effect can only be observed at ultra-low temperatures ($\lesssim 1$ K), i.e. much lower than the materials' magnetic transition temperatures ($\sim$10 K) and the achieved local exchange gap sizes ($\sim 100$ K)[9]. Recently, it was discovered that stoichiometric $MnBi_2Te_4$ (MBT) is an intrinsic magnetic topological insulator in which Mn atoms are embedded as uniform layers in the crystal structure, forming a natural heterostructure[10-18]. The resulting exchange gap is predicted to be larger than in magnetically doped TIs and thus supposed to be more resilient to inhomogeneities[12, 14]. Similar to $Bi_2Te_3$, the charge carrier concentration in MBT can be tuned by Sb substitution, resulting in $MnBi_{2-x}Sb_xTe_4$ (MBST), with charge neutrality achieved at $x \approx 0.64$ (Refs.[19, 20]). As a result, the compensated compound $MnBi_{1.36}Sb_{0.64}Te_4$ is a promising candidate to overcome technical limitations of magnetically doped BST. However, the experimentally observed surface state gaps in both MBT and MBST are varying widely from tens of meV all the way down to zero[14, 19-23]. A likely reason for this discrepancy is the effect of intrinsic defects, such as vacancies and anti-sites, which give rise to local variations in the electronic structure. However, the roles of the different defect types are not yet fully understood. In this letter, we elucidate the influence of intrinsic defects on the local electronic structure of compensated $MnBi_{1.36}Sb_{0.64}Te_4$ bulk crystals by scanning tunnelling microscopy/spectroscopy (STM/S). While we find that substitutional Sb atoms are incorporated uniformly and are an effective way to tune the Fermi level, the inhomogeneous distribution of $(Bi,Sb)_{Mn}$ anti-site defects gives rise to significant doping fluctuations, which close the transport gap.

## Results

Topographic images of the MBST surface (Figs. 1a and b) display atomically flat surfaces of hexagonal atomic structure with lattice constant $a = 4.3$ Å corresponding of the topmost Te layer[24]. Different kinds of defects are evident on the surface and can be identified based on the crystal structure and earlier reports on pure $MnBi_2Te_4$ (Refs.[25, 26]), as shown schematically in Fig. 1c. In particular, $Mn_{(Bi/Sb)}$ anti-sites are the dominant type of defect observed on the sample surface, which result in dark triangular features centred on the Bi/Sb atoms in the second layer (Fig. 1b). These features are similar to previous STM studies of Mn-doped $Bi_2Te_3$, where Mn atoms replace Bi atoms in the second layer, resulting in the three covalently bonded Te atoms above the substitutional atom to show a decreased topographic height[27]. We further observe $(Bi/Sb)_{Te}$ anti-sites sparsely, which replace a chalcogen atom in the topmost layer[25, 28] and appear as bright circular spots on the sample surface (Fig. 1a). $(Bi,Sb)_{Mn}$ anti-sites are another type of defect reported frequently in the literature. However, because of their location further below the surface, in the fourth atomic layer, they result in a smaller contrast in the STM topography[25], which makes them hard to distinguish, especially if the defect concentration is high. Furthermore, we find that the locations of Bi and Sb atoms cannot be distinguished because their similar valence electron structures do not give rise to a significant STM contrast.

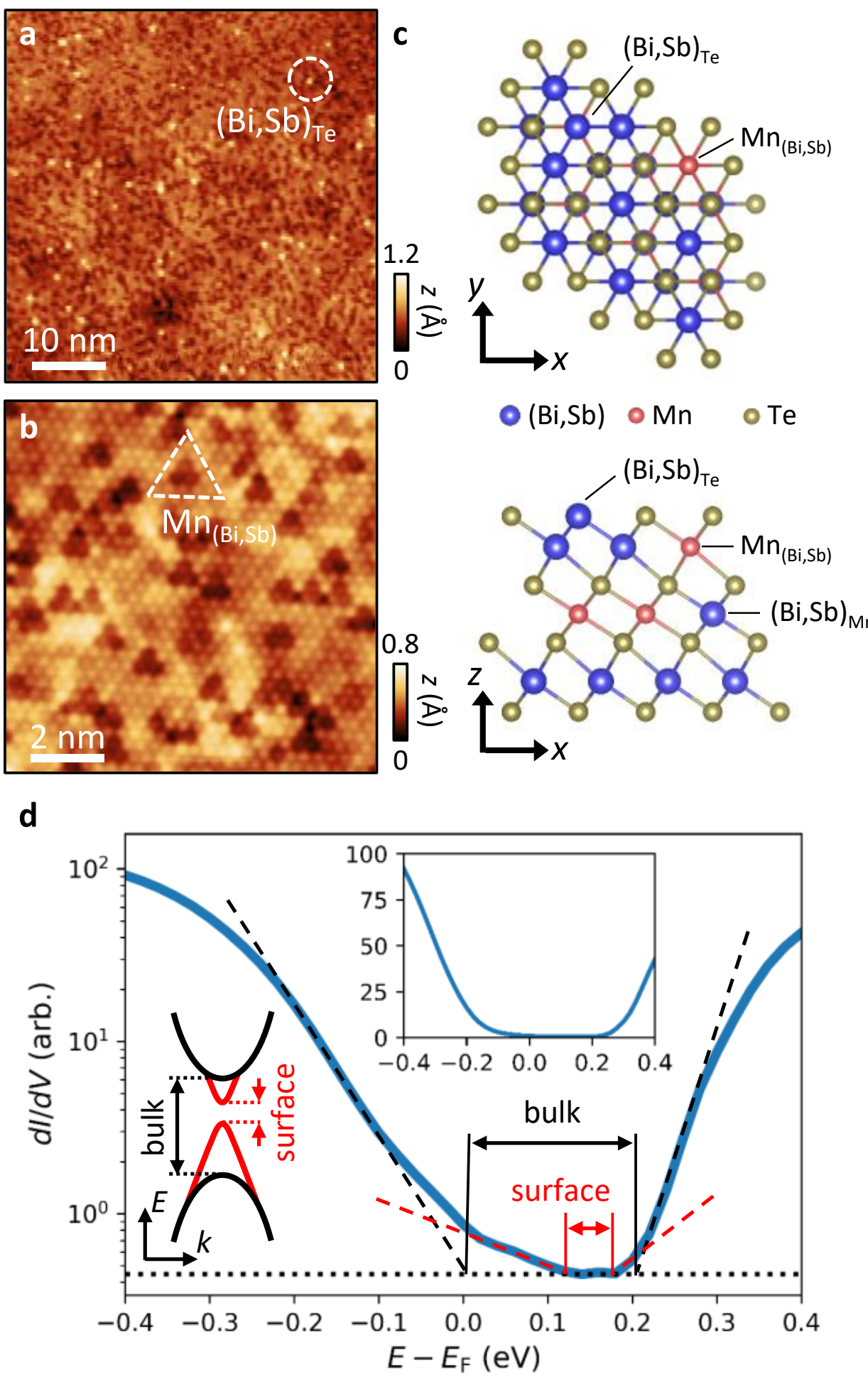

**Figure 1. Surface defects and electronic spectrum of MBST.** (a, b) STM images showing intrinsic defects in large scale and atomic resolution scans. (c) Atomic models of typical intrinsic defects in MBST viewed from the top (top panel) and side (bottom panel). (d) Typical tunnelling spectrum of MBST averaged over an area of $50 \times 50\ \text{nm}^2$, plotted on a logarithmic scale, as well as a linear scale (upper inset). The flat region in the inset corresponds to the bulk band gap. In the log-scale plot, the surface state inside the bulk band gap is evident and shows another gap – the surface state exchange gap. A corresponding band schematic is shown as the left inset. We estimate the bulk band gap ($\approx 200$ meV) and surface state exchange gap ($2\Delta \approx 56$ meV) from linear extrapolation of the band edges (black and red dashed lines, respectively) to the measurement noise level (horizontal dotted line), similar to Refs.[29, 30].

The concentrations of $Mn_{(Bi/Sb)}$ and $(Bi/Sb)_{Te}$ anti-sites are estimated from topographic images to be 9.1% and 0.46%, respectively. These defect concentrations are higher than in pure $MnBi_2Te_4$[31] but lower than in pure $MnSb_2Te_4$[32] grown by the same group. This observation is in agreement with theoretical calculations that showed a lower $Mn_{(Bi/Sb)}$ defect formation energy in $MnSb_2Te_4$ than in $MnBi_2Te_4$[33]. Figure 1d shows a spatially averaged tunnelling spectrum of the MBST sample surface, which is in excellent agreement with a recent photoemission study of similar MBST crystals[34]. We can readily identify the bulk band gap, which is comparable to that of MBT[19, 30]. However, in contrast to pure MBT, which is typically electron-doped with the Fermi energy $E_\mathrm{F}$ in the conduction band, in our MBST sample $E_\mathrm{F}$ is located at the top of the bulk valence band. Inside the bulk band gap, we observe the gapped surface state, which is described by the dispersion of a massive Dirac cone as $E_\pm(k) = E_\mathrm{D} \pm \sqrt{(\hbar v_\mathrm{F} k)^2 + \Delta^2}$, where $v_\mathrm{F}$ is the Fermi velocity, $E_\mathrm{D}$ is the Dirac point energy, i.e. the doping level, and $\Delta$ is the magnetic exchange gap[9]. The corresponding surface state valence and conduction band edges are $E_\mathrm{VB} = E_\mathrm{D} - \Delta$ and $E_\mathrm{CB} = E_\mathrm{D} + \Delta$, respectively.

In the large area topographic images (Fig. 1a) we also observe long-range height variations indicative of local doping fluctuations. Such variations can not only be caused by the defects evident on the sample surface but also by subsurface defects, such as $(\mathrm{Bi/Sb})_\mathrm{Mn}$ anti-sites[25] and the random distribution of the Sb dopants. Because Sb dopants and $\mathrm{Mn}_{(\mathrm{Bi/Sb})}$ anti-sites both induce hole doping, while $(\mathrm{Bi/Sb})_\mathrm{Mn}$ induces electron doping[25], a revelation of the exact roles for each dopant requires further investigations. However, in contrast to pure MBT, where defect concentrations can be low enough to enable identification of isolated defects, partial Sb substitution leads to significant defect densities, which results in an overlap of the effects of neighbouring defects. This problem is aggravated by the extended nature of $(\mathrm{Bi/Sb})_\mathrm{Mn}$ anti-site features in STM, stemming from their location three layers below the sample surface. As a result, we find that atomic-scale variations in the electronic properties are small compared to the long-range variations that result from regions of different concentrations of particular defect types (see Supplementary Note 1). To resolve this issue, we use a statistical approach to analyse the influence of the defects onto the electronic properties of MBST in the following.

## Statistical analysis of the electronic properties

To assess $\Delta$ and $E_\mathrm{D}$ experimentally, we record $dI/dV$ maps in a smaller energy range, corresponding to the surface state, where we find significant variations in the individual spectra (Figs. 2a, b). At each pixel of the $dI/dV$ map, we extract the valence band onset $E_\mathrm{VB}(x, y)$ and conduction band onset $E_\mathrm{CB}(x, y)$ as the energy at which the $dI/dV$ signal drops below a fixed threshold value (see Methods section for details). The resulting maps are shown in Figs. 2c, d. It is evident that there are regions that are more hole-doped (red) and electron-doped (blue) in similar locations in $E_\mathrm{VB}(x, y)$ and $E_\mathrm{CB}(x, y)$. When plotting the histogram of Figs. 2c and d, we find that both band edges exhibit variations in the range of $\sim 50$ meV (Fig. 2e). By fitting Gaussian distributions $\sim \exp\left(-\frac{(E-\tilde{E})^2}{2\sigma^2}\right)$, we find the average positions of the valence and conduction band edges to be $\tilde{E}_\mathrm{VB} = \tilde{E} \pm \sigma = (116 \pm 24)$ meV and $\tilde{E}_\mathrm{CB} = (172 \pm 20)$ meV, respectively. Because the histograms are almost perfectly normal distributed, the fit errors for $\tilde{E}_\mathrm{VB}$ and $\tilde{E}_\mathrm{CB}$ are only ~0.06 meV. On average, the two band edges have a separation of $\tilde{E}_{CB} - \tilde{E}_{VB} = 2\tilde{\Delta} \approx 56$ meV with respect to each other, which is consistent with the exchange gap observed in the averaged tunnelling spectra (Figs. 1d and 2b). The fact that there is an overlap between the conduction and valence band distributions in Fig. 2e necessitates the presence of charge puddles, effectively closing the transport gap[35]. To further analyse if this band overlap results from a local closing of the gap or comes from the varying doping level across the surface, we extract the local gap size $\Delta(x,y) = (E_{\mathrm{CB}}(x,y) - E_{\mathrm{VB}}(x,y))/2$ and present the map in Fig. 2f, where we find similar spatial variations as for the band edges. A Gaussian fit to the histogram of the extracted gap sizes (Fig. 2g) results in an average exchange gap of $\tilde{\Delta} = 27.80$ meV, with fit error 0.06 meV. The standard deviation of the gap is $\sigma_{\Delta} = 9.70$ meV, with a vanishing number of data points for which the gap is fully closed. This observation indicates no significant presence of regions with such high defect density that the surface gap collapses[36]. The smaller standard deviation of the gap size ($\sim$10 meV) compared to the propagated standard deviation of the band edges ($\sqrt{(20\ \mathrm{meV})^2 + (24\ \mathrm{meV})^2}/2 \approx 16$ meV) indicates that the observed variations in Figs. 2c, d are predominantly due to different local doping, which does not primarily alter the exchange gap size. By performing a Fourier transformation $\Delta(x,y) \rightarrow \hat{\Delta}(k_x,k_y)$ we find that the exchange gap is uniform in $k$-space (Fig. 2h). This is in contrast to magnetically doped BST in which the gap resembles a warped Dirac cone with hexagonal symmetry[9].

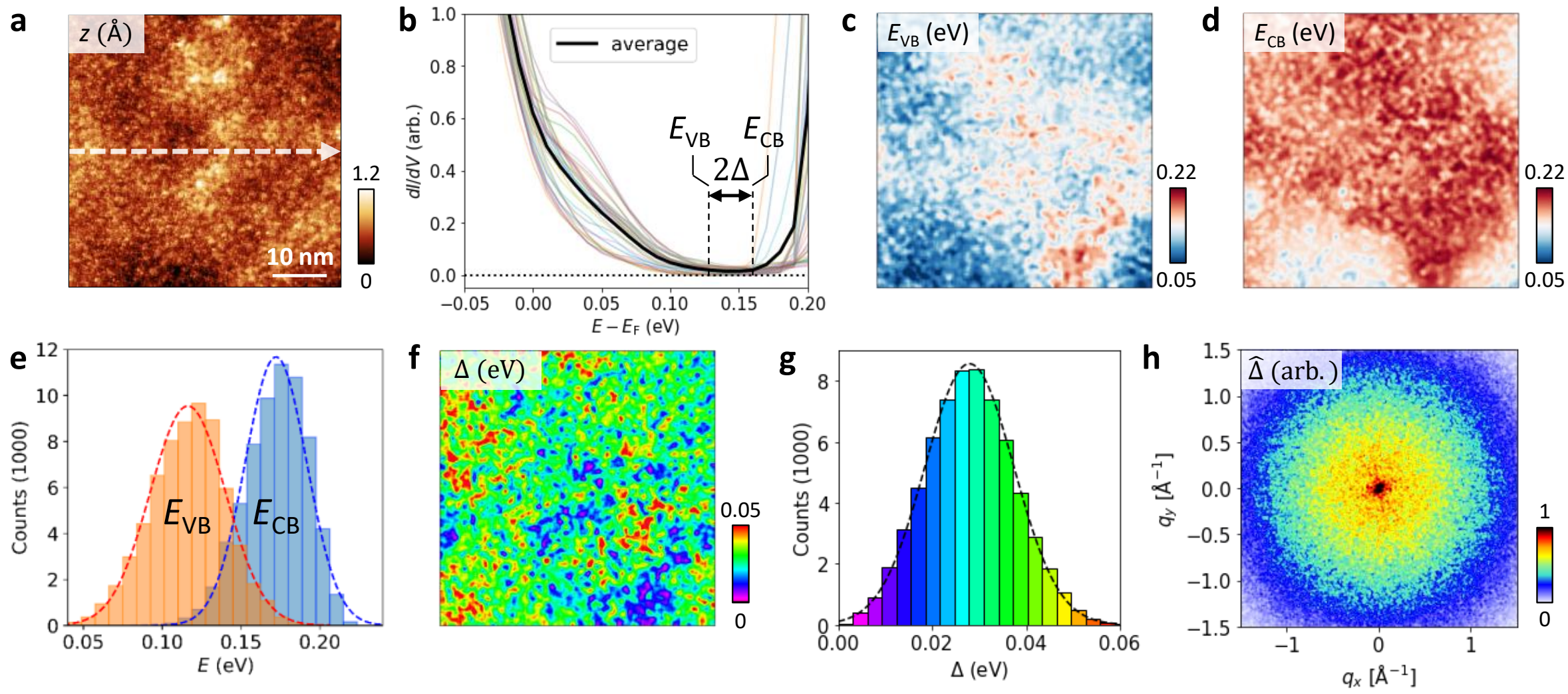

**Figure 2. Extraction of the surface state doping level and gap size.** (a) Large-area STM topography showing long-range fluctuations on the sample surface. (b) Variation of the surface state spectra along the line indicated in (a) and their average. The band edge positions $E_{\mathrm{VB}}$ and $E_{\mathrm{CB}}$ are indicated. (c, d) Map of the variation of $E_{\mathrm{VB}}$ and $E_{\mathrm{CB}}$ in the same area as shown in (a). (e) Histogram of the data shown in (c, d). Gaussian fits to the peaks result in band edge positions $\tilde{E}_{\mathrm{VB}} = (116 \pm 24)$ meV and $\tilde{E}_{\mathrm{CB}} = (172 \pm 20)$ meV. (f) Map of the extracted surface state gap $\Delta = (E_{\mathrm{CB}} - E_{\mathrm{VB}})/2$. (g) Histogram of (f) with Gaussian fit resulting in $\tilde{\Delta} = (28 \pm 10)$ meV. (h) Fast Fourier transformation of (f) showing a uniform gap in $k$-space. Color bar units are given in the panel titles.

To test the hypothesis of mostly rigid band shifts by Sb doping, we analyse the dependence of the gap size on the band edge positions for each data pixel. When plotting $E_{\mathrm{CB}}$ as function of $E_{\mathrm{VB}}$ (Fig. 3a), we find that the data is described well by the equation $E_{\mathrm{CB}} = E_{\mathrm{VB}} + 56\ \mathrm{meV} = E_{\mathrm{VB}} + 2\tilde{\Delta}$. This linear dependence of $E_{\mathrm{CB}}$ on $E_{\mathrm{VB}}$ with a slope of unity indicates mostly rigid band shifts, which is expected for local changes in the charge doping of the crystal, such as with varying amounts of Sb doping. We also observe a group of data points deviating from this overall trend (highlighted with red ellipse in Fig. 3a), which can be ascribed to an effect of $(Bi/Sb)_{Mn}$ defects, as discussed below. Plotting $\Delta$ as function of the Dirac point energy $E_D = (E_{VB} + E_{CB})/2$, results in a relatively small correlation coefficient ($r = -0.255$), which indicates a disorder-induced variation of the gap independent of the doping level. Nevertheless, an overall trend is discernible, which can be described by a linear fit $\Delta(E_D) = -0.061 \cdot E_D + 23$ meV (Fig. 3b). This trend of the gap size can be explained by an increase in the number of $Mn_{(Bi/Sb)}$ defects with increasing Sb content, which decreases the surface gap due to ferrimagnetic ordering[19, 21, 32]. Although the gap size in the observed range of $E_D$ varies by ~40%, it has overall weak dependence on the doping level, allowing us to conclude that Sb doping realizes an effective tuning of $E_F$ near $x = 0.64$. From the observed trend, we expect that it will be possible to tune $E_F$ to mid-gap while simultaneously increasing the gap size by slightly decreasing the Sb content. At mid-gap condition, the predicted average gap size corresponds to the offset of the fit ($\Delta = 23$ meV). We note that for even lower Sb concentration ($x \leq 0.2$), it was reported that the gap increases with increasing Sb content[37]. In combination with our results, this could indicate that there is an optimum composition in the range $0.2 < x < 0.64$ for which the surface gap is maximized.

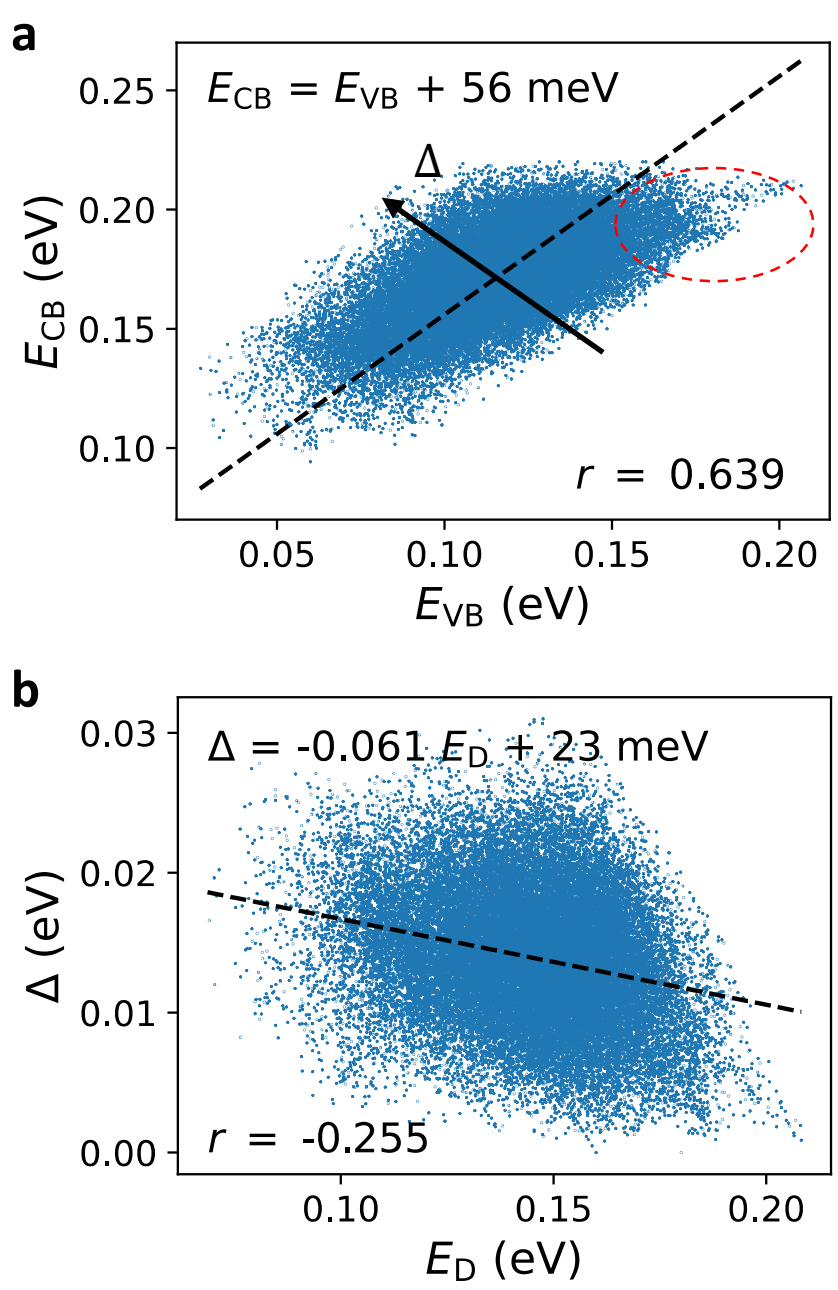

**Figure 3. Statistical analysis of doping effect and surface state gap.** (a) $E_{CB}$ as function of $E_{VB}$ for each pixel of Figs. 2c, d. The correlation coefficient of the data is $r = 0.639$. The indicated line of $E_{CB} = E_{VB} + 56$ meV corresponds to isoelectronic doping and describes the data well. The red ellipse highlights a cloud of outliers resulting from $(Bi/Sb)_{Mn}$ defects. The arrow indicates the axis of increasing surface state gap $\Delta$. (b) The surface state gap as function of the Dirac point energy $E_D = (E_{VB} + E_{CB})/2$ shows only a small slope and little correlation ($r = -0.255$), indicating a weak dependence of the two quantities.

## Quasiparticle interference

To analyse the effect of disorder in more detail, we study the electronic surface structure of the MBST by quasiparticle interference (QPI) measurements. QPI patterns are calculated from the *dI/dV* maps by Fourier transformation $dI/dV(x, y, E) \rightarrow \widehat{dI/dV}(q_x, q_y, E)$, where $q_x, q_y$ are aligned with high symmetry axes of the atomic lattice. Figures 4a, b show the *dI/dV* maps and QPI patterns, at energies corresponding to the bulk valence band ($E = -150$ meV), the surface state just below the exchange gap ($E = 50$ meV) and the bulk conduction band ($E = 220$ meV), respectively. We note that the *dI/dV* intensity fluctuations at the Dirac cone energy and the conduction band energy are qualitatively roughly opposite to each other and show a similar pattern as the doping fluctuations evident in Figs. 2c, d. In the QPI pattern of the bulk bands, we see broad features, owing to the dispersion of the corresponding bands[19, 20, 26]. In contrast, the surface state QPI pattern shows a smaller Dirac cone signature that is similar to non-magnetic BST[38-40]. However, the Dirac cone warping observed in our QPI data is smaller than in BST, in agreement with photoemission experiments on MBST[20]. In Fig. 4c we plot the quasiparticle dispersion $E(q)$ by taking cross sections of the QPI patterns at each energy along the $\bar{\Gamma} - \bar{\mathrm{M}}$ and $\bar{\Gamma} - \bar{\mathrm{K}}$ directions and stacking them on top of each other. The bulk band gap and gapped Dirac cone of the surface state is visible in the energy range $E = [-50\ \mathrm{meV}, 160\ \mathrm{meV}]$ (indicated as red dashed lines in Fig. 4c). Extracting the Dirac cone group velocity from its QPI dispersion, we find that it agrees well with photoemission spectroscopy of MBST (Refs.[41]). Furthermore, the QPI intensity of the surface state is suppressed in the energy range $E = [100\ \mathrm{meV},\ 160\ \mathrm{meV}]$, in agreement with the presence of the exchange gap[19].

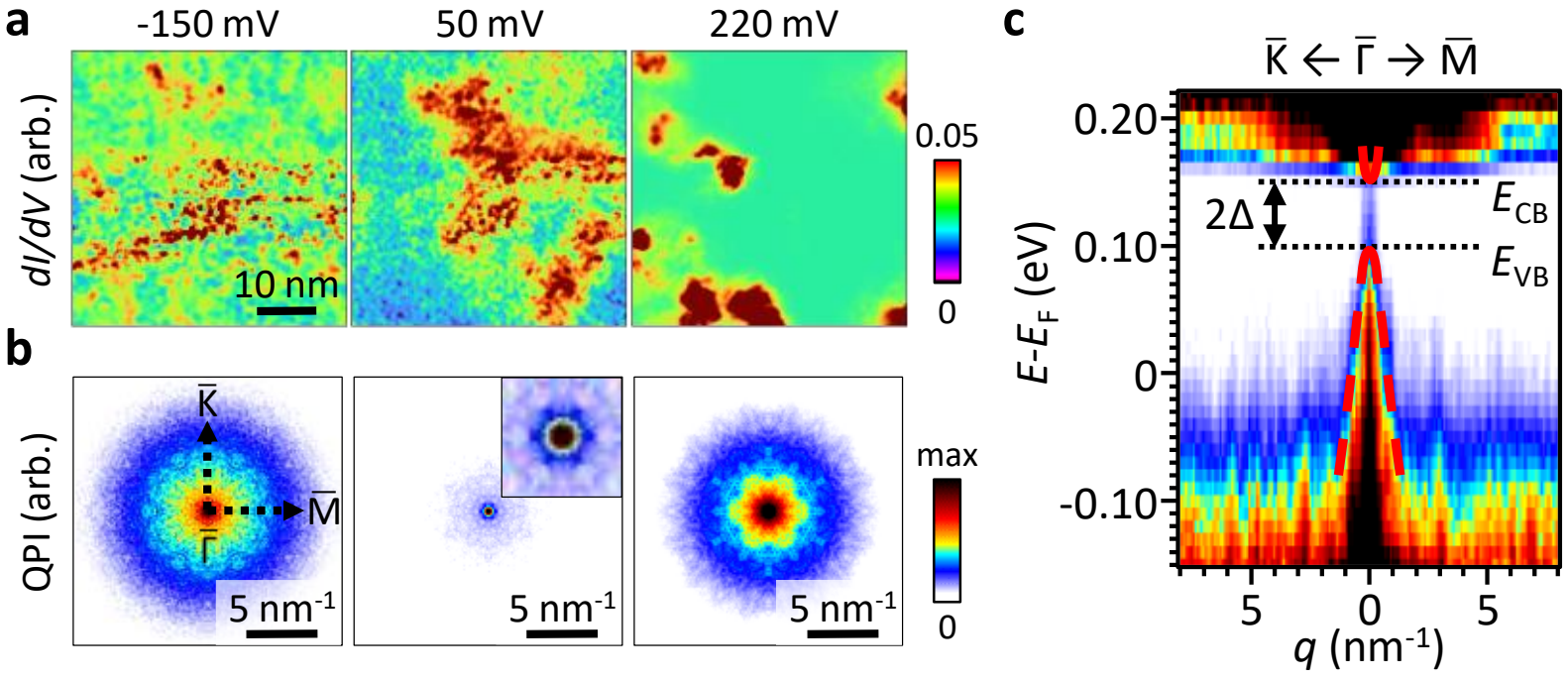

**Figure 4. Quasiparticle interference mapping.** (a) *dI/dV* maps measured in the surface region shown in Fig. 2a at three energies, corresponding to bulk valence band ($E = -150$ meV), bulk conduction band ($E = 220$ meV) and surface state ($E = 50$ meV). (b) QPI maps calculated from (a). The middle panel inset is a zoom into the surface state, showing the warped Dirac cone. (c) Quasiparticle dispersion resulting from stacking of QPI maps with indicated gapped Dirac cone (red dashed lines) and its gap (black dotted lines).

To analyse the effect of the defects on the QPI patterns, we scrutinize regions of different local doping by classifying the *dI/dV* map into two subsets where $E_\mathrm{D} < 150$ meV and $E_\mathrm{D} \geq 150$ meV, respectively, and analyzing them separately. We perform the same QPI analysis as described above for each of the two regions (red and blue areas in Fig. 5a). This approach is viable because it has been shown that even sparsely sampled *dI/dV* maps are sufficient to calculate representative QPI patterns[42]. We call this technique subset-QPI (S-QPI) in the following. Figures 5b and c show the S-QPI dispersions and the averaged spectra for each region, respectively. We find that the dispersion of the two regions are overall similar, but shifted in energy by $(20 \pm 7)$meV with respect to each other, owing to the more electron-doped character of region A compared to region B. In the S-QPI dispersion of region A, we further observe an additional feature in $\bar{\Gamma} - \bar{\mathrm{M}}$ direction at $q \approx 3\ \mathrm{nm}^{-1}$, which is also clearly reflected in the corresponding average spectrum as a shoulder at $E \approx 180$ meV (red arrow in Fig. 5b). This defect state is known to result from subsurface $(\mathrm{Bi/Sb})_{\mathrm{Mn}}$ defects[25] and we thus attribute additional counts in this energy range in Figs. 2e and 3a to this defect state. In contrast, this defect state is not observed in region B, and consequently the S-QPI displays a more ideal dispersion of the gapped Dirac cone. We conclude that the more electron-doped region A has a significantly higher concentration of $(\mathrm{Bi/Sb})_{\mathrm{Mn}}$ defects than region B, which is in line with the reported local electron doping resulting from such defects[25].

## Disentangling the defect roles

To further analyse the details of the surface state gap variations as function of $E_{\mathrm{VB}}$ and $E_{\mathrm{CB}}$, we extract the slope and intercept of the linear fit shown in Fig. 3a with respect to the indicated $\Delta$-axis, resulting in $\Delta(E_{\mathrm{VB}}) = -0.239 \cdot E_{\mathrm{VB}} + 0.056$ meV, and $\Delta(E_{\mathrm{CB}}) = 0.109 \cdot E_{\mathrm{CB}} + 0.09$ meV, respectively. As such, the data indicates that the gap tends to close from the valence band side because of the higher slope there. This observation is in agreement with the combination of the $(\mathrm{Bi/Sb})_{\mathrm{Mn}}$ defect state being located at the conduction band edge, at $E \approx 180$ meV, and the electron doping caused by these defects. As a result, these two effects conspire to decrease the transport gap in the vicinity of high $(\mathrm{Bi/Sb})_{\mathrm{Mn}}$ defect concentrations, ultimately leading to the manifestation of charge puddles, when the Fermi energy is tuned into the surface state gap. However, such regions only make up a small fraction of the total number of data points (highlighted in Fig. 3a) and we find that the average gap sizes of region A and region B differ by $(3.9 \pm 0.2)$ meV while the average doping difference is $(24.4 \pm 0.9)$ meV (Figs. 5b, c and Supplementary Note 2), consistent with our results above. Since the surface state gap in MBT was shown to critically depend on the local concentration of $\mathrm{Mn}_{(\mathrm{Bi/Sb})}$ defects[21], the concentration of this type of defect is expected to be comparable in region A and B, i.e. independent of the $(\mathrm{Bi/Sb})_{\mathrm{Mn}}$ defect distribution. We confirm this behaviour by directly estimating the defect concentration in the two regions from topography maps, where we find 9.25% and 9.02% $\mathrm{Mn}_{(\mathrm{Bi/Sb})}$ defects for region A and B, respectively (see Supplementary Note 3). In addition, because the concentration of $\mathrm{Mn}_{(\mathrm{Bi/Sb})}$ defects depends on the Sb content of the crystal[32], we expect the Sb concentration in the two regions to be similar as well. As a result, the observed variation in the doping level of the two regions must result from the different $(\mathrm{Bi/Sb})_{\mathrm{Mn}}$ defect concentrations (see Supplementary Note 3 for further discussion).

Lastly, we observe that the overall intensity of the Dirac cone in the S-QPI data of region A (Fig. 5b) is lower than in region B (Fig. 5c). Since it was shown that the surface state depth profile critically depends on the defect concentration[21, 43, 44], we speculate that the higher concentration of $(\mathrm{Bi/Sb})_{\mathrm{Mn}}$ defects in region A pushes the surface state into the interior of the crystal and away from its surface. As a result, the surface state contributes less to the tunnelling current, which depends exponentially on the tip-surface state distance. This effect could also explain the lower intensity of the surface state compared to pure MBT (see Ref. [19] and Supplementary Note 4), as well as earlier reports of a 'buried' surface state[26].

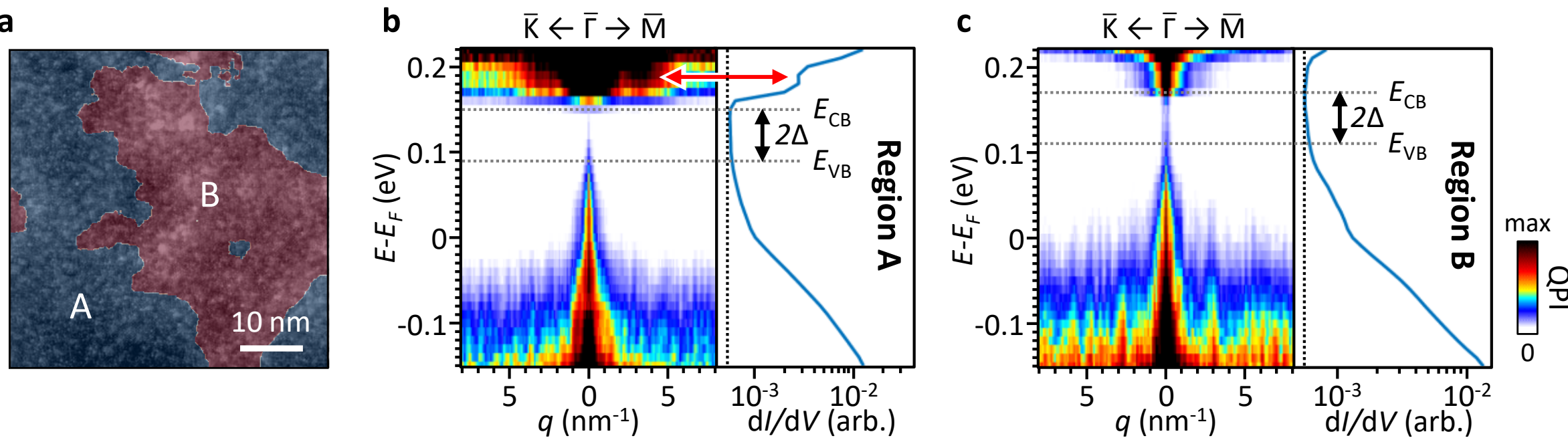

**Figure 5. Quasiparticle dispersions of regions of different doping level.** (a) Classification of areas that are more electron-doped (region A, blue, $E_{\mathrm{D}} < 150$ meV) or hole-doped (region B, red $E_{\mathrm{D}} \geq 150$ meV) superimposed on the topography from Fig. 2a. (b, c) Quasiparticle dispersions and average spectra of regions A and B, respectively. Horizontal dotted lines indicate the surface state gaps, respectively.

# Conclusion

We conclude that substitutional Sb atoms are incorporated uniformly in MBST and are an effective way to tune the Fermi level. While the effects of the individual defects in MBST have overall comparable effects as in pure MBT, the interaction of the different defect types with the Sb substitution creates a more complicated picture. In detail, the surface state exchange gap is strongly influenced by anti-site defects, where $(\mathrm{Bi,Sb})_{\mathrm{Mn}}$ defects give rise to local electron doping, which is the main source of doping fluctuations in this compound, and $\mathrm{Mn}_{(\mathrm{Bi/Sb})}$ defects give rise to overall hole doping and a decrease of the local surface state gap. We find the positions of $(\mathrm{Bi,Sb})_{\mathrm{Mn}}$ defects to be independent of the $\mathrm{Mn}_{(\mathrm{Bi/Sb})}$ defects, and the doping fluctuations caused by accumulations of $(\mathrm{Bi,Sb})_{\mathrm{Mn}}$ defects to gives rise to the formation of charge puddles, which close the transport gap. As such, we are able to disentangle the roles of the individual types of defects on the electronic properties of MBST. Going forward, the minimization of fluctuations in the doping and surface state gap, through engineering of $(\mathrm{Bi,Sb})_{\mathrm{Mn}}$ and $\mathrm{Mn}_{(\mathrm{Bi/Sb})}$ defects, respectively, will be a critical task to achieve a robust transport gap in this promising material platform.

# Methods

## Scanning Tunnelling Microscopy/Spectroscopy

The STM/S experiments were performed in two ultra-high vacuum (UHV) systems (base pressure $\sim 10^{-10}$ mbar) hosting commercial low-temperature scanning tunnelling microscopes (Scienta Omicron LT-Nanoprobe and LT-STM) both operating at liquid helium temperatures (~5 K). PtIr and W tips were obtained by electromechanical etching. PtIr tips we further sharpened by focus ion beam milling. All tips were prepared by indentation into Au(111) surfaces, where the Shockley surface state was used as a benchmark for a clean tip spectrum. *dI/dV* spectra were collected using standard lock-in techniques at oscillation amplitudes of $V_{\mathrm{AC}} = 3$ mV to 5 mV. The spectroscopy in Fig. 2 and 5 is performed on a $256 \times 256$ px$^2$ grid. Scan parameters: Figure 1a: $V_{\mathrm{sample}} = (E - E_{\mathrm{F}})/\mathrm{e} = -1$ V, $I_{\mathrm{t}} = 10$ pA; Figure 1b: $V_{\mathrm{sample}} =$ $-0.8$ V, $I_t = 100$ pA; Figure 2: $V_{sample} = -0.15$ V, $I_t = 150$ pA. The crystal structures in Fig. 1 were created using *VESTA*[45] and topography data was visualized using *Gwyddion*[46].

To detect the gap size and doping levels (Fig. 2), we use an algorithm, which consists of the following steps: 1) Spatial averaging of the raw data with a $5 \times 5$ px$^2$ ($0.98 \times 0.98$ nm$^2$) window using Gaussian weighting; 2) In each spectrum: Subtraction of a flat background corresponding to the minimum value of the spectrum. 3) Detection of $E_{CB}$ and $E_{VB}$ as the voltages at which the data crosses the *dI/dV* threshold value of 0.05 (arb. units). 4) Another round of spatial averaging as described in step 1) to mitigate outliers of the band edge detection.

We have tested different averaging window sizes around $1 \times 1$ nm$^2$ and found little influence on the quantitative results of our analysis. We avoid gliding average smoothing of the individual curves due to possible boundary artefacts, which could affect the detected gap sizes, as the conduction band edges are in proximity to the upper limit of the spectra. A similar algorithm has been used in Ref. [47].

The thermal broadening of the tunnelling spectra ($5\ k_B T \approx 2$ meV) becomes comparable to our experimentally observed gaps ($2\Delta$) only for a tiny fraction ($< 1\%$) of the data (see Fig. 2g). When thermal broadening becomes significant, our detection algorithm will tend to underestimate the gap. However, we expect this effect to be small compared to the *dI/dV* noise.

**Sample preparation**

MBST was grown by a flux method, resulting in single crystals of dimension 3×4 mm and a thickness of 0.1–1 mm[24]. The stoichiometry of the crystals was determined using energy-dispersive x-ray analysis, which did not show any compositional variation as function of position on the crystal or between crystals from the same batch. The crystals are cleaved in the UHV chamber at room temperature and pressures of $\sim 10^{-10}$ mbar just before the introduction to the STM stage for characterization. Measurements on multiple cleaves and positions on the sample show consistent crystal quality (Supplementary Note 4). Photoemission and transport measurements of samples of similar composition have shown a transition temperature of $T_N \sim 24$ K and non-closing of the surface state gap through the magnetic transition[24, 34].

## Data availability

The data supporting the findings of this study are available from the authors upon reasonable request.

## Code availability

The code used to analyse the data is available from the authors upon reasonable request.

## Acknowledgements

We acknowledge the assistance of James Burns and Jonathan Poplawsky for focused ion beam milling of the STM tips. We acknowledge Mao-Hua Du for helpful discussions. This work was supported by Center for Nanophase Materials Sciences (CNMS), which is a U.S. Department of Energy (DOE), Office of Science User Facility at Oak Ridge National Laboratory. J. Y acknowledges support from the U.S. DOE, Office of Science, Basic Energy Sciences, Materials

Sciences and Engineering Division. A.-P.L. acknowledges support from the U.S. DOE, Office of Science, National Quantum Information Science Research Centers, Quantum Science Center. F.L. acknowledges funding from the Alexander von Humboldt foundation through a Feodor Lynen postdoctoral fellowship, the Deutsche Forschungsgemeinschaft (DFG, German Research Foundation) within the Priority Programme SPP 2244 (project no. 443416235) and the Bavarian Ministry of Economic Affairs, Regional Development and Energy within Bavaria's High-Tech Agenda Project "Bausteine für das Quantencomputing auf Basis topologischer Materialien mit experimentellen und theoretischen Ansätzen". M.K. acknowledges support from the U.S. DOE, Office of Science, Basic Energy Sciences, Materials Science and Engineering Division at the Ames National Laboratory, which is operated for the U.S. DOE by Iowa State University under Contract No. DE-AC02-07CH11358.

## Author Contributions

F.L., M.K., W.K., H.C. and P.V. performed the experiments. F.L., M.K. and W.K. analysed the data. J.Y. grew the samples. A.-P.L. and H.H.W. supervised the experiments. F.L., M.K., W.K and A.-P.L. wrote the manuscript. All authors have given approval to the final version of the manuscript.

## Competing interests

The authors declare no competing interests.